\documentclass[conference]{IEEEtran}
\IEEEoverridecommandlockouts
\usepackage{amsmath,amssymb,amsfonts}
\usepackage{algorithmic}
\usepackage{graphicx}
\usepackage{subcaption}
\usepackage{booktabs}
\usepackage{multirow}
\usepackage{textcomp}
\usepackage{xcolor}
\usepackage{hyperref}
\usepackage{pifont}
\usepackage{soul}
\usepackage{adjustbox}
\hypersetup{
     colorlinks=true,
     citecolor=blue,
}
\usepackage[
backend=biber,
style=ieee,
maxnames=6, minnames=6,
doi=false,isbn=false,url=false,eprint=false
]{biblatex}
\def\BibTeX{{\rm B\kern-.05em{\sc i\kern-.025em b}\kern-.08em
    T\kern-.1667em\lower.7ex\hbox{E}\kern-.125emX}}
\begin{document}

\title{Serenade: A Singing Style Conversion Framework Based On Audio Infilling
% \thanks{Identify applicable funding agency here. If Unused, delete this.}
}

\author{\IEEEauthorblockN{Lester Phillip Violeta}
\IEEEauthorblockA{\textit{Graduate School of Informatics} \\
\textit{Nagoya University}\\
Nagoya, Japan}
\and
\IEEEauthorblockN{Wen-Chin Huang}
\IEEEauthorblockA{\textit{Graduate School of Informatics} \\
\textit{Nagoya University}\\
Nagoya, Japan}
\and
\IEEEauthorblockN{Tomoki Toda}
\IEEEauthorblockA{\textit{Information Technology Center} \\
\textit{Nagoya University}\\
Nagoya, Japan}
}

\maketitle
% ContentVec linguistic features, MIDI, loudness, and the masked target mel-spectrogram
\begin{abstract}
We propose Serenade, a novel framework for the singing style conversion (SSC) task. Although singer identity conversion has made great strides in the previous years, converting the singing style of a singer has been an unexplored research area. We find three main challenges in SSC: modeling the target style, disentangling source style, and retaining the source melody. To model the target singing style, we use an audio infilling task by predicting a masked segment of the target mel-spectrogram with a flow-matching model using the complement of the masked target mel-spectrogram along with disentangled acoustic features. On the other hand, to disentangle the source singing style, we use a cyclic training approach, where we use synthetic converted samples as source inputs and reconstruct the original source mel-spectrogram as a target. Finally, to retain the source melody better, we investigate a post-processing module using a source-filter-based vocoder and resynthesize the converted waveforms using the original F0 patterns. Our results showed that the Serenade framework can handle generalized SSC tasks with the best overall similarity score, especially in modeling breathy and mixed singing styles. We also found that resynthesizing with the original F0 patterns alleviated out-of-tune singing and improved naturalness, but found a slight tradeoff in similarity due to not changing the F0 patterns into the target style.
\end{abstract}

\begin{IEEEkeywords}
singing style conversion, audio infilling, cyclic training, vocoder post-processing
\end{IEEEkeywords}

\section{Introduction}
Voice conversion (VC) \cite{vcc2016, vcc2018, vcc2020}, the task known as changing the speaker information while keeping linguistic information unchanged, has undergone rapid changes in the age of generative AI. Several VC approaches were first widely compared in the Voice Conversion Challenge (VCC) 2016 \cite{vcc2016}. More difficult tasks such as non-parallel conversion and cross-lingual VC were explored in VCC 2018 \cite{vcc2018} and VCC 2020 \cite{vcc2020}. In VCC 2020, results showed that top systems could already synthesize at a naturalness and similarity score to ground-truth recordings. The Singing Voice Conversion Challenge (SVCC) 2023 then focused on singing voice conversion \cite{huang2023svcc}, where top systems using end-to-end frameworks had statistically similar naturalness scores to ground truth recordings, but had a large gap from the ground truth in similarity scores. Upon analyzing the results, we found that evaluating ground truth recordings was also difficult, as the singing styles significantly affected the perception of identity of the singer.  %Thus, modern neural-based VC systems have been shown to successfully convert the timbre of a source waveform into that of a prompt speaker or singer at a high quality. 

Thus, as singing voices contain more sophisticated and specific features than singer identity, controlling the singing style has more practical applications and need to be investigated to advance the field. Previously, SingStyle111 \cite{singstyle111} was released with various singing style labels, opening the potential for this task. A large-scale and open-sourced dataset called GTSinger \cite{zhang2024gtsinger} was also released, containing singing data with various parallel song phrases sung in different singing styles. With the release of GTSinger, new VC subapplications such as singing style conversion (SSC) can now be explored. As these datasets are new, SSC has been a relatively unexplored task.

In this work, we investigate the novel SSC task and propose a framework called Serenade to accomplish this. We identified three primary challenges in completing this task. First, we needed to develop a robust framework capable of accurately modeling the singing style of a reference singer. Second, we had to successfully disentangle the source singer's singing style to ensure high-quality synthesis. Finally, aside from capturing the singing style, we needed to preserve the original melody and synthesize the waveform in the correct notes to avoid out-of-tune singing.

In particular, we resolved these problems by adopting the audio infilling task from text-to-speech (TTS) \cite{borsos22_speechpainter, le2023_voicebox, e2tts, liu2024speechflow} to generate a mel-spectrogram in the target singing style. To further disentangle the singing style from the source audio, we also show the use of cyclic training \cite{CycleGAN2017, tobing19_cyclevae}. Finally, to alleviate out-of-tune singing, we introduce a post-processing scheme which resynthesizes the converted waveforms in the correct notes, using a source-filter vocoder SiFiGAN \cite{sifigan}.

\begin{figure*}[t!]
  \centering
  \includegraphics[width=18cm]{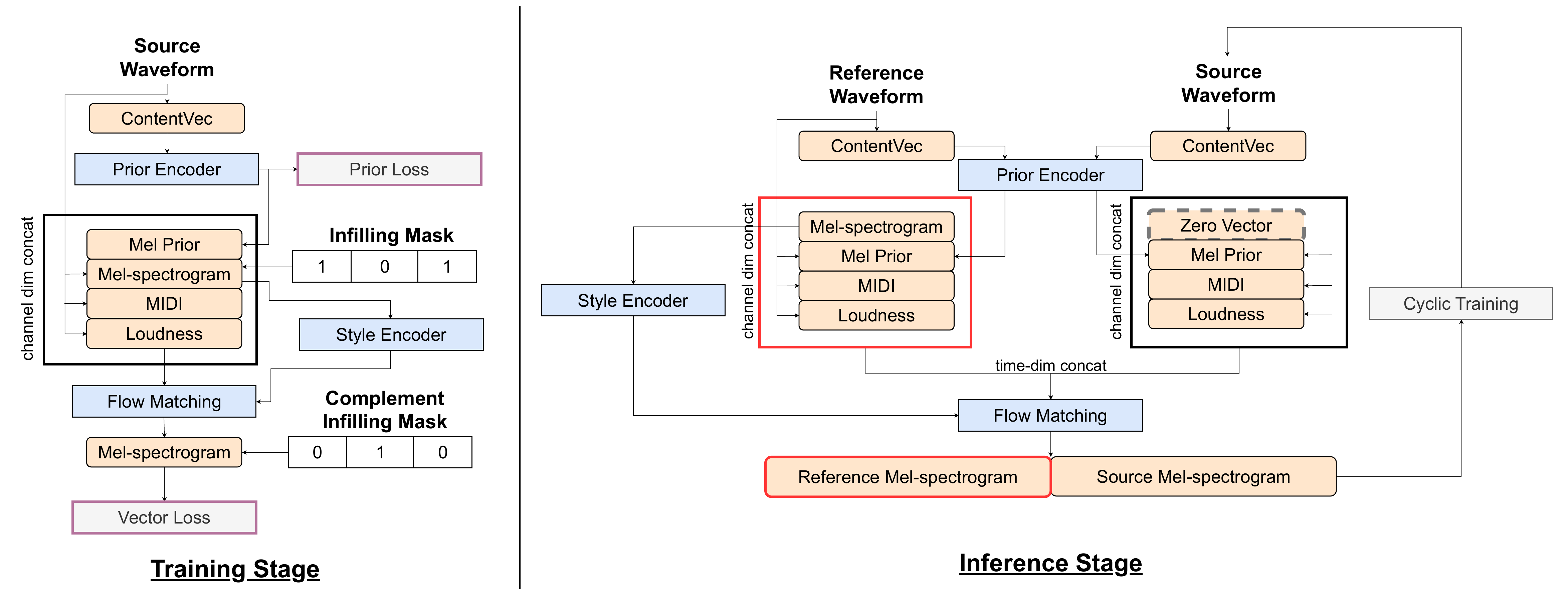} % 8cm
  \caption{An overview of the Serenade architecture. The left hand side shows the training stage, which reconstructs the masked segments of an input utterance. The right hand side shows the inference stage, which using reference and source waveforms, transforms the singing style of the source into the reference. Once the mel-spectrogram is generated, the frames from the reference waveform (indicated in red) are discarded, and the waveform is synthesized using a HiFiGAN vocoder, and further postprocessed using a SiFiGAN vocoder.}
  \label{fig:overall}
\end{figure*}

The contributions of this work are as follows:
\begin{itemize}
\item We propose a novel framework called Serenade to accomplish the SSC task. We adopt the audio infilling task and predict the masked segment of the mel-spectrogram with a flow-matching model using its complement as conditioning features. We find that adopting the audio infilling task allows us to model more singing styles.
\item To disentangle the singing style from the source waveform, we use cyclic training \cite{CycleGAN2017, tobing19_cyclevae}. Converted waveforms from an initial model are synthesized and used as conditioning features to reconstruct its original source waveform. This makes the framework more robust by learning how to disentangle more source styles.
\item Finally, we retain the melody of the source waveform better by using a source-filter vocoder SiFiGAN \cite{sifigan} to post-process the converted waveforms. This is done by analyzing the acoustic features of the converted waveforms and resynthesizing these using the source waveform's F0 pattern. We analyze the tradeoff this causes in naturalness and similarity.
\item We open-source the code \footnote{Code repository: \url{https://github.com/lesterphillip/serenade}} and pretrained models \footnote{Demo page: \url{https://lesterphillip.github.io/serenade_demo}} to invite researchers to build on this work and explore SSC further.
\end{itemize}

\section{Related Work}
\subsection{Singing voice conversion}
Singing voice conversion (SVC) takes in singing waveforms as inputs and generates identity-converted singing waveforms. SVCC 2023 \cite{huang2023svcc} showed that state-of-the-art systems can generate highly natural speech, where top systems \cite{ning2023svccwhisper} utilized end-to-end waveform generation systems based on variational autoencoders. However, current SVC frameworks are only limited to converting the global identity of a target speaker, and are not fully capable of copying singing styles from the reference waveform. Thus, SSC which converts the singing style of an audio is an underexplored task.

\subsection{Audio infilling task for TTS}
\label{sec:related_infilling}
The audio infilling task has recently become a popular framework in the TTS field. Initial works such as SpeechPainter \cite{borsos22_speechpainter} showed the ability to infill gaps in speech and copy prosody and speaker information from the surrounding audio and text conditioning features, but was limited to a mask size of a single second. Subsequent works build on this framework \cite{le2023_voicebox} and generalize the TTS framework through a audio infilling task, where segments of the target mel-spectrogram are masked and used as conditioning information, and the masked segments are used as targets to regularize the network. The key point is the use of the flow-matching model \cite{lipman2023flow}, which compared to SpeechPainter, allows the network to generate masked segments longer than a single second. Recent work such as E2 TTS \cite{e2tts} proposed an inference framework for the audio infilling task, where a reference utterance's text and mel-spectrogram, along with the source text, can be used as conditioning information to generate the target mel-spectrogram. Thus, several previous works have shown that the audio infilling framework can successfully copy style from a reference utterance.

\section{Proposed Method}
We propose Serenade, a novel framework that converts the singing style of a singer. The framework addresses the three main challenges in this task: style modeling, source style disentanglement, and melody retention. We provide a visualization of the framework in Fig. \ref{fig:overall} and detail how each challenge is resolved with the framework.

% \subsection{Audio infilling task for SSC}
% To model the singing style from a reference singer, we adopt ideas from the audio infilling papers as discussed in Section \ref{sec:related_infilling}, but adapt it for the SSC task.
\subsection{Optimal transport conditional flow-matching}\label{AA}
We first describe the flow-matching model used to generate the masked segments of the target mel-spectrogram. We adopt the conditional flow-matching with optimal transport (OT-CFM) objective proposed in \cite{lipman2023flow}. %The proposed formulation is like any generative model, wherein it transforms a simple initial Gaussian distribution $x_0$ into a complex target distribution $x_1$ (in this case, the mel-spectrograms). However, the key novelty of this framework lies in the transformation process, which leverages optimal transport to construct a time-dependent vector field $u_t(x|x_1)$, for $t \in [0, 1]$ using a neural network $v_t(x;\theta)$ with parameters $\theta$. \
The transformation process, which leverages optimal transport to construct a time-dependent vector field $u_t(x|x_1)$, for $t \in [0, 1]$ using a neural network $v_t(x;\theta)$ with parameters $\theta$, goes from a simple initial Gaussian distribution $x_0$ into a complex target distribution $x_1$ (in this case, the mel-spectrograms). Once trained, the network defines a flow $\phi_{t}$ that transforms the initial distribution $x_0$ into $x_1$. Using the OT-CFM objective \cite{lipman2023flow} of $\phi^{\mathrm{OT}}_t(x_0) = (1 - (1-\sigma_{\mathrm{min}})t)x_0 + t x_1$, we can use the time derivative of this path as a target, $u^{\mathrm{OT}}_t(\phi^{\mathrm{OT}}_t(x_0)\vert x_1) = x_1-(1-\sigma_{\mathrm{min}})x_0$, to optimize the neural network parameters, where $\sigma_{\mathrm{min}}$ is a small positive constant for stability.

\subsection{Training scheme: Audio Infilling}
The audio infilling task training is conducted as follows and is visualized in the left hand side of Fig. \ref{fig:overall}. Given target mel-spectrogram features $\hat{y}\in\mathbb{R}^{D\times T}$ (where $D$ is the feature dimension and $T$ is the sequence length), we create a mask ${m}\in\mathbb{R}^{1\times T}$ where a segment from $t_m$ to $t_n$ is set to zeros. The mask ${m}$ is applied to $\hat{y}$ to create the target feature $\hat{y}_m$. Another mask ${m_c}\in\mathbb{R}^{1\times T}$ is created from the complement of $m$, and is applied to $\hat{y}$ to make a conditioning feature $\hat{y_c}$. In addition to this, other conditioning features $\hat{c}\in\mathbb{R}^{D\times T}$ are used, which are not masked. It might be noted that in TTS, these conditioning features are usually not aligned with the target features, and an alignment module \cite{le2023_voicebox} or filler tokens \cite{e2tts} are used to match the sequence lengths. However, in the context of SSC, the conditioning features $\hat{c}$ are also extracted from the audio, which means that no alignment methods are necessary.

We choose features that individually represent the linguistic, melody, and energy features of the singing voice. We use ContentVec \cite{2022qiancontentvec} to represent linguistic features, which has been proven better than other linguistic features in SVC tasks \cite{yamamoto2023svcc}. We also adopt the encoder prior loss proposed in \cite{popov2021gradtts} to generate a mel-spectrogram prior. We observed that introducing this prior loss function is necessary to allow the training to converge. To represent melody information, the pitch features need to be normalized as the pitch patterns also contain information of the singing style. We choose MIDI features to represent melody as it is a commonly used feature in singing synthesis. In our experiments, we also compare the representations that are directly extracted from the audio waveforms using a neural network \cite{2023yongphonememidi} (which we found to be more correlated to the input audio) and the ground truth score MIDI labels provided in the dataset (which we found to be more disentangled from singing style) to identify the ideal MIDI representation. For loudness features, we found that using them as is performed the best. Finally, a global style encoder (GST) \cite{wang2018gst} extracts a time-independent style feature from $\hat{y}$ and is used as another conditioning feature.

\subsection{Inference scheme}
During inference, we follow the same procedure proposed in E2 TTS \cite{e2tts}. As seen in the right hand side of Fig. \ref{fig:overall}, we take in two singing waveforms: the reference (which contains the target singing style) and the source (which is the waveform to be converted). The acoustic features are from both reference and source waveforms. However, only the mel-spectrogram from the reference waveform is used, while the mel-spectrogram from the source waveform is replaced with a zero vector of the same shape. The features are concatenated together in the time-dimension, which simulates the masking procedure done during training. Once the target mel-spectrogram is generated, we discard the frames corresponding to the reference mel-spectrogram. We used the first-order Euler forward ODE-solver to generate the mel-spectrogram given the initial distribution $x_0$. We use a HiFiGAN \cite{kong2020hifigan} neural vocoder to generate the initial waveforms from the mel-spectrogram.

\subsection{Fine-tuning scheme: Cyclic Training}
As the audio infilling task is trained on a reconstruction loss, we find that the model is incapable of completely disentangling the source singing style, leading to low-quality synthesis. To remedy this, we adopt cyclic training \cite{CycleGAN2017, tobing19_cyclevae} to further regularize the model with a style conversion objective. We first train a model using the audio infilling task described above, then create synthetic data by converting the training set in other singing styles. We fine-tune the model further by using the conditioning features $\hat{c}\in\mathbb{R}^{D\times T}$ extracted from the synthetic data of converted singing styles but its original source mel-spectrogram as both the masked conditioning feature $\hat{y_c}$ and target feature $\hat{y}_m$. 

\subsection{Post-processing scheme: SiFiGAN vocoder}
Despite the use of large-scale data and cyclic training, we observed that the converted samples did not retain the melody very well, resulting in out-of-tune singing. To resolve this, we adopt a post-processing scheme with SiFiGAN \cite{sifigan}, a source-filter vocoder, which could modify the pitch patterns of a waveform in high-quality. SiFiGAN follows the signal analysis process of WORLD vocoder \cite{world} by extracting the fundamental frequency (F0), mel cepstrum (mcep), and band aperiodicity (bap) features of a waveform, and synthesizes the waveform using the neural network SiFiGAN.

We use this idea to resynthesize the converted waveforms using the source waveform's pitch patterns, therefore correcting the out-of-tune singing patterns. Given the analyzed WORLD features of a source waveform $\langle \text{F0}_{\text{src}}, \text{mcep}_{\text{src}}, \text{bap}_{\text{src}} \rangle$ and a converted waveform $\langle \text{F0}_{\text{cvt}}, \text{mcep}_{\text{cvt}}, \text{bap}_{\text{cvt}} \rangle$, we instead use $\langle \text{F0}_{\text{src}}, \text{mcep}_{\text{cvt}}, \text{bap}_{\text{cvt}} \rangle$ to resynthesize the converted waveforms. We also linearly shift $\text{F0}_{\text{src}}$ to the statistics of the reference waveform using a mean-variance shift \cite{yamamoto2023svcc}. However, although this method can easily correct out-of-tune singing, keeping the source F0 patterns can have demerits in converting the singing style. We further investigate this tradeoff in the results.

% \begin{table*}[t]
\begin{table*}[htbp]
    \centering
    \footnotesize
    \caption{Subjective evaluation results of N-MOS and SIM (also in each target singing style) with 95\% confidence score. Higher scores indicate better performance. Scores in bold and underlined are the best overall system, while scores values in bold signify systems that have no statistically significant difference to the best system based on Wilcoxon signed-rank tests.}
    \label{tab:subjective}
    \centering
    \begin{adjustbox}{width=1\textwidth,center}
    \begin{tabular}{lcccccccccc}
    \toprule
    \multicolumn{1}{c}{\textbf{System}} & 
    \multicolumn{1}{c}{\textbf{\begin{tabular}[c]{@{}c@{}}MIDI\\ Type\end{tabular}}} & 
    \multicolumn{1}{c}{\textbf{\begin{tabular}[c]{@{}c@{}}Large-scale\\ Training\end{tabular}}} & 
    \multicolumn{1}{c}{\textbf{\begin{tabular}[c]{@{}c@{}}Cyclic\\ Training\end{tabular}}} & 
    \multicolumn{1}{c}{\textbf{\begin{tabular}[c]{@{}c@{}}SiFiGAN\\ Post-processing\end{tabular}}} & & & \multicolumn{4}{c}{\textbf{SIM (Target Styles)}}\\
    \cmidrule(lr){8-11}
    % \textbf{System} & \textbf{MOS} & \textbf{SIM} & \textbf{Breathy} & \textbf{Falsetto} & \textbf{Mixed} & \textbf{Pharyngeal} \\
    & & & & & \textbf{N-MOS} & \textbf{SIM} & \textbf{Breathy} & \textbf{Falsetto} & \textbf{Mixed} & \textbf{Pharyngeal} \\
    \toprule
    Ablation 1 & Score & Omitted & Omitted & Omitted & 1.78 $\pm$ 0.06 & 2.11 $\pm$ 0.07 & 2.34 $\pm$ 0.15 & 2.08 $\pm$ 0.14 & 2.07 $\pm$ 0.13 & 1.94 $\pm$ 0.13 \\
    Ablation 2 & Audio & Omitted & Omitted & Omitted & 2.05 $\pm$ 0.06 & 1.96 $\pm$ 0.07 & 2.10 $\pm$ 0.15 & 1.98 $\pm$ 0.14 & 1.80 $\pm$ 0.12 & 1.95 $\pm$ 0.14 \\
    Ablation 3 & Audio & Applied & Omitted & Omitted & 1.98 $\pm$ 0.06 & 2.48 $\pm$ 0.07 & 2.60 $\pm$ 0.15 & 2.37 $\pm$ 0.15 & 2.42 $\pm$ 0.12 & 2.55 $\pm$ 0.16 \\
    Serenade & Audio & Applied & Applied & Omitted & 2.17 $\pm$ 0.06 & \ul{\textbf{2.60 $\pm$ 0.07}} & \ul{\textbf{2.76 $\pm$ 0.14}} & 2.46 $\pm$ 0.14 & \ul{\textbf{2.54 $\pm$ 0.14}} & \textbf{2.64 $\pm$ 0.14} \\
    Serenade SiFiGAN & Audio & Applied & Applied & Applied & 3.06 $\pm$ 0.06 & \textbf{2.56 $\pm$ 0.08} & \textbf{2.64 $\pm$ 0.15} & \textbf{2.61 $\pm$ 0.16} & 2.27 $\pm$ 0.15 & \textbf{2.71 $\pm$ 0.16} \\
    \midrule
    NU-SVC \cite{yamamoto2023svcc} & - & - & - & - & \ul{\textbf{3.30 $\pm$ 0.07}} & 2.48 $\pm$ 0.08 & 1.98 $\pm$ 0.14 & \ul{\textbf{2.68 $\pm$ 0.17}} & \textbf{2.46 $\pm$ 0.15} & \ul{\textbf{2.78 $\pm$ 0.14}} \\
    GT & - & - & - & - & 4.03 $\pm$ 0.06 & 3.11 $\pm$ 0.07 & 3.22 $\pm$ 0.13 & 3.19 $\pm$ 0.12 & 2.93 $\pm$ 0.14 & 3.08 $\pm$ 0.14 \\
    \bottomrule
    \end{tabular}
    \end{adjustbox}
\end{table*}

\section{Experimental Setting}
\subsection{Datasets}
We used the GTSinger \cite{zhang2024gtsinger} dataset, which is a semi-parallel dataset containing songs sung in different styles. We focus on four singing styles, Breathy, Falsetto, Mixed Voice, and Pharyngeal. The entire dataset contains around 80 hours of data. We resample all audio to 24 kHz sampling rate.

We selected two songs to be used for the dev and test set, both of which have songs parallel in all four singing styles. Note that different from the original task which converts from the control style into another target style, we did not convert from the control style and instead convert from each style to another. Since the test set song contains 12 phrases in four styles, there were a total of 144 converted utterances after converting each phrase into each of the three other styles. The reference utterances were randomly selected from the train set and the same reference waveform of each singing style is used throughout the inference.

In addition to GTSinger, we used the large-scale singing datasets used in the NU-SVC system \cite{yamamoto2023svcc} which consists of around 120 hours of data. We also conducted further data augmentation of GTSinger by pitch shifting the training data. We used SiFiGAN \cite{sifigan} and shifted the pitch of the waveforms from -4 to 4 semitones.

\subsection{Model architecture}
For the prior encoder and GST style encoder, we used the same model configurations as in NU-SVC \cite{yamamoto2023svcc}. For the UNet used in the flow-matching model, we base it on the 1D UNet architecture proposed in Matcha-TTS \cite{mehta2024matcha} for high-quality synthesis. However, to further improve style modeling, we also conditioned the style embeddings after every residual block and increased the attention head dimension to 256 and the channel size to 512. 

\subsection{Evaluation tests and baseline system}
For subjective evaluations, we recruited 106 listeners and conducted the 5-scale mean opinion score test evaluating naturalness (N-MOS). Following the SVCC 2023 protocol, the definition of naturalness was on the listeners' discretion. Also following the SVCC 2023 evaluation, we also conducted a 4-scale XAB singing style similarity test, by showing listeners the source (A) and target (B) phrases and rate which was more similar to the converted phrase (X). Choosing B results in a score of 4, while choosing A results in a score of 1. X, A, and B are parallel and thus contain the same lyrics but with just different styles. In cases where the listener has to evaluate the ground truth's similarity, we instead use a pitch-shifted version (of either half an octave up or down) of the source and target waveforms to ensure that listeners would not be easily able to distinguish the samples to be completely similar. 

For the baseline system, we use NU-SVC \cite{yamamoto2023svcc}, a DDPM-based SVC system, but replace the DDPM decoder with the flow-matching decoder. NU-SVC is similar to Serenade, but does not use the audio infilling task, and directly predicts the entire target mel-spectrogram from the conditioning features. For ablation studies, we investigate the use of audio and score MIDI, the use of large-scale pretraining, the use of the cyclic training, and the use of the SiFiGAN post-processing.

\section{Results and Discussion}
\subsection{Effectiveness of Serenade's audio infilling task}
We show that the proposed Serenade SiFiGAN framework had the highest overall N-MOS in the proposed systems. Despite the use of cyclic training to improve similarity scores, the N-MOS in Serenade was still very low due to the out-of-tune singing, possibly due to inaccuracies during MIDI estimation. However, Serenade SiFiGAN helped alleviate this by using the original F0 and correcting out-of-tune singing, resulting in a much higher naturalness score. We also observed that there was a slight degradation in naturalness in Serenade SiFiGAN compared to NU-SVC. This is a downside of the SiFiGAN postprocessing scheme, as simply replacing the F0 with the source's would have likely introduced small mismatches in WORLD features.

Next, we analyze the similarity scores in detail. We first observed that evaluating the singing style may be a difficult task in itself, as ground truth scores compared to its pitch shifted versions resulted in just an average of score of around 3 out of 4. Moreover, although Serenade SiFiGAN had significantly improved naturalness, we observed that the post-processing caused a slight degradation in similarity scores but nevertheless has no statistically significant difference with Serenade. Next, we find that Serenade systems beat NU-SVC overall primarily because of its ability to model the breath features better. We see that Serenade can greatly model breathy singing style, and is slightly better in modeling mixed singing style (which also contain breath features), showing that NU-SVC has difficulties in modeling breath features. However, NU-SVC and Serenade SiFiGAN were better in modeling falsetto and pharyngeal styles. Thus, disentangling the source F0 is more important in modeling singing styles such as breathy and mixed, while the linear pitch shift during inference is more important in modeling singing styles such as falsetto and pharyngeal. We conclude that the necessity of converting the F0 patterns is dependent on the target singing style.

\subsection{Ablation studies}
We first investigate compared the use of score MIDI labels provided in GTSinger (Ablation 1) with the use of the audio-extracted MIDI using the pretrained network \cite{2023yongphonememidi} (Ablation 2). We found that using audio-extracted MIDI was better than using score MIDI, primarily because we opted to use audio-extracted MIDI during inference to keep the system practical for SSC. Nevertheless, we still found that using audio-extracted MIDI, and being more correlated to the input waveform, was more helpful in normalizing the pitch.

As we validated the use of audio-extracted MIDI, using a large-scale dataset and data augmentation (Ablation 3) is now considered viable as score labels are not necessary. We found that using large-scale data can improve the similarity scores. Unexpectedly, the N-MOS values did not improve compared to Ablations 1 and 2. We conclude that the limit in improvement was due to the model being incapable of synthesizing the correct F0 patterns from just MIDI information, and that other pitch normalization strategies should be explored in the future.

\subsection{Correlations between objective and subjective tests}
Finding a well-correlated objective metric can help future researchers validate ideas before running expensive subjective tests. With no previous works on SSC, we found that reference-based metric FAD \cite{gui2024fad} using MERT \cite{li2024mert} embeddings was the most correlated to the similarity scores, with a 0.93 correlation score. Nevertheless, large-scale evaluations with a multiple variety of systems still need to be conducted to find the best objective metric.

\section{Conclusions}
We proposed Serenade, a novel framework to accomplish the SSC task and resolved the three main issues we found in this task, namely: singing style modeling, source style disentanglement, and melody retention. We showed a more generalized SSC framework and better modeling in breath features due to the audio infilling. Next, we used a cyclic training strategy to further disentangle the singing style of the source singer. Finally, we were able to alleviate out-of-tune singing by introducing a SiFiGAN vocoder post-processing method. We conclude that the Serenade framework showed better similarity scores in breathy and mixed singing styles than the baseline system but with a slight degradation in naturalness to do so.

\section*{Acknowledgment}
This work was partly supported by JST CREST under Grant Number JPMJCR19A3 and by JST AIP Acceleration Research JPMJCR25U5, Japan.
% \section*{References}
\printbibliography
\vspace{12pt}

\end{document}